# Surfactant effects in monodisperse magnetite nanoparticles of controlled size

P. Guardia[a,c], B. Batlle-Brugal[a], A. G. Roca[b], O. Iglesias[a], M. P. Morales[b], C. J. Serna[b], A. Labarta[a] and X. Batlle[a,*]

[a]*Departament de Física Fonamental and Institut de Nanociencia i Nanotecnologia (IN2UB), Universitat de Barcelona, Martí i Franquès 1, 08028- Barcelona, Catalonia, Spain*
[b]*Instituto de Ciencia de Materiales de Madrid, CSIC, Sor Juana Inés de la Cruz 3, Cantoblanco 28049, Madrid, Spain*
[c]*Institut Català de Nanotecnologia, Campus UAB, 08193, Bellaterra, Catalonia, Spain*

**Abstract**

Monodisperse magnetite $Fe_3O_4$ nanoparticles of controlled size within 6 and 20 nm in diameter were synthesized by thermal decomposition of an iron organic precursor in an organic medium. Particles were coated with oleic acid. For all samples studied, saturation magnetization $M_s$ reaches the expected value for bulk magnetite, in contrast to results in small particle systems for which $M_s$ is usually much smaller due to surface spin disorder. The coercive field for the 6 nm particles is also similar to that of bulk magnetite. Both results suggest that the oleic acid molecules covalently bonded to the nanoparticle surface yield a strong reduction in the surface spin disorder. However, although the saturated state may be similar, the approach to saturation is different and, in particular, the high-field differential susceptibility is one order of magnitude larger than in bulk materials. The relevance of these results in biomedical applications is discussed.

## 1. Introduction

Magnetic nanoparticles [1] have attracted much research over the recent years due to their potential interests in a variety of biomedical applications [2]. They posses an increasing relevance as diagnostic and therapeutic tools, such as, for example, contrast agents in magnetic resonance imaging, drug delivery to tumour cells and cancer treatment by hyperthermia, and cell separation and purification, among others. Magnetic nanoparticles for biomedical applications should comply with a variety of requirements, including: (i) superparamagnetic behaviour at room temperature, in order to avoid particle aggregation; (ii) large saturation magnetisation, so as to show a large response under the application of a magnetic field; (iii) a limiting size in the order of 20 nm for *in vivo* applications, and (iv) bio-compatibility, such that nanoparticles are usually coated with either biological or bio-compatible molecules.

Magnetic nanoparticles are also an ideal system to study finite-size and surface effects, all them yielding new phenomena and enhanced properties with respect to their bulk counterpart [1].

In this paper, uniform magnetite $Fe_3O_4$ nanoparticles within 6 and 20 nm in diameter were synthesised by thermal decomposition at high temperature of an iron organic precursor in an organic medium [3]. Surprisingly enough, particles coated with oleic acid show the saturation magnetisation, $M_s$, expected for bulk magnetite, for all the sizes studied, as already suggested by Roca *et al.* [4], in contrast to typical results for nanoparticles [5], which do not show $M_s$ of bulk $Fe_3O_4$ up to about 150 nm in size. However, although the saturated state in oleic acid-coated magnetite nanoparticles may be close to the bulk one, the approach to saturation is different and, in particular, the high-field differential susceptibility (slope of the hysteresis loop at high magnetic fields) is at least one order of magnitude larger than in bulk materials. The aim of this work is to analyse this behaviour.

## 2. Experimental

* Corresponding author. Tel.: +34-934021172; fax: +34-934021149.
 *E-mail address*: xavier@ffn.ub.es (X.Batlle).

Magnetite nanoparticles were synthesised departing from different precursors, either Fe(acac)$_3$ or Fe(CO)$_5$. Oleic acid was used as surfactant in all cases, the Fe: oleic acid ratio was kept at 1:3 and the organic solvent was chosen in each case so as to maximise size and shape uniformity [2-4]. Three samples are discussed in this paper, with the following average diameters obtained from transmission electron microscopy (TEM): 5.7 ± 1.1 nm, 10.4 ± 1.0 nm and 16.7 ± 3.0 nm. The particles are very uniform in size (Figure 1), with polydispersion below 20%, and consist of a mixture of diamond-, cubic- and triangular-shaped particles [4]. X-ray diffraction (XRD) patterns show that all samples consist of highly crystalline particles with an inverse spinel structure and lattice parameters similar to those of magnetite. Average diameters obtained from XRD for those three samples are 5.8 ± 1.3 nm, 14.2 ± 1.9 nm and 19.9 ± 2.4 nm. Infra-red (IR) spectra and thermogravimetric analyses (TGA) evidence that oleic acid molecules are bonded covalently to the particle surface.

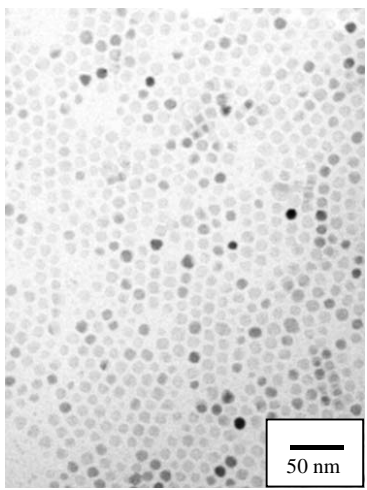

Fig. 1. TEM micrograph of magnetite nanoparticles with an average diameter of 10.4 ± 1.0 nm

## 3. Results and discussion

Figure 2 shows the zero-field-cooling (ZFC) and field-cooling (FC) magnetization, measured at 50 Oe. All samples were prepared in the form of powders after evaporating the solvent, which does not affect the oleic acid coating, as demonstrated by IR and TGA. For the case of 6 nm particles, the ZFC evidences a narrow size distribution (rapid increase of the signal up to temperature of the maximum of the ZFC, $T_{max}$ = 28 K, and Curie-Weiss-like behaviour above $T_{max}$). The progressive increase of the FC curve for temperatures below $T_{max}$ indicates the absence of relevant dipolar interactions, suggesting that they are largely reduced due to the surfactant taking the particles apart (Figure 1). However, the fact that for this sample the ZFC and FC curves joint well above $T_{max}$ suggests that some aggregates are yet present. For samples with 10 and 17 nm particles, $T_{max}$ evolves with particle size, and as the average particle magnetization is much larger than for the 6 nm case, the separation among particles is not enough to reduce the dipolar interactions. The FC curves are thus flat below $T_{max}$. We note that for the 17 nm sample, $T_{max}$ is already well above room temperature, while the increase in the FC curve above about 100 K for the 10 nm sample is due to the fact that the cooling process was probably too fast, so that some of the particles remain in a metastable state until the system is heated again and the signal increases, as shown in many fine particle systems [1].

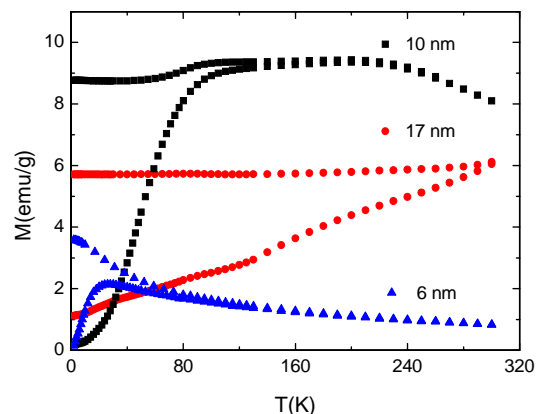

Fig. 2. ZFC/FC curves for magnetite nanoparticles surfacted with oleic acid, as a function of particle size. The cooling field was 50 Oe.

Figure 3 displays the hysteresis loops at 5 K for the three samples (6, 10 and 17 nm in particle size). In the high magnetic field region, they can be fitted to $M(H) = M_s + \chi_d H$, being $M_s$ the zero-field saturation magnetization and $\chi_d$ the high-field differential susceptibility that accounts for the surface spin disorder [6-7]. $M_s$ and $\chi_d$ are obtained from the average of the H → ± $H_{max}$ branches, yielding 79(1) emu/g, 81.8(1) emu/g and 84(4) emu/g, respectively, which are roughly the saturation magnetization of the bulk material ($M_s^{bulk}$ = 82 emu/g) [8]. This is contrast with the reported values for Fe$_3$O$_4$ nanoparticles ($M_s$ ≈ 50 emu/g for 4 nm particles), for which the bulk value was not reached up to 150 nm [5]. This result suggests that the crystal field associated with the new O$^{2-}$ surface ligans of the oleic acid coating that the Fe surface cations suffer, resembles that of the Fe bulk cations and helps to reduce the surface spin disorder.

However, although the saturation state of the nanoparticles in this paper and bulk magnetite may be similar, the approach to that state is different. For example, $\chi_d$ values are 4(2)10$^{-5}$ emu/g, 2.6(2)10$^{-5}$ emu/g, and 6(3)10$^{-5}$ emu/g, respectively, for the 6, 10 and 17 nm samples, while the typical value for bulk particles is in the order of 10$^{-6}$ emu/g [2,6]. This suggests that the oleic acid coating does not erase completely the surface spin disorder. Even in this situation, we note that $\chi_d$ for those 4 nm particles above [5] was 1.2·10$^{-4}$ emu/g.

The inset to Figure 3 shows the low field region of the hysteresis loop. The coercive fields $H_c$ are 175(10) Oe, 280(10) and 370(10) Oe, for the 6 nm, 10 nm and 17 nm samples, respectively. The value for the 6 nm particles is in agreement with the expected for bulk magnetite, $H_c^{bulk}$, taking into account that for a random distribution of non-interacting magnetite nanoparticles $H_c^{bulk} = 0.64 K_v/M_s^{bulk}$ = 175-210 Oe, being $K_v$ = 1.1-1.3·10$^5$ erg/cm$^3$ the first magnetocrystalline anisotropy of bulk magnetite [5,9]. This result points out again the relevance of the oleic acid coating in reducing the surface spin disorder, since the effective anisotropy constant $K_{eff}$ for this sample must be close to that of the bulk counterpart. This is in contrast with the usual results in nanoparticles, for which $K_{eff}$ is much larger than the bulk value [2,7,10], due to the surface contribution, $K_s$, through the expression $K_{eff} = K_v + (S/V)K_s$, being S and V the particle surface and volume, respectively [2]. Even more, according to the later expression, the effective anisotropy constant of nanoparticles decreases with increasing particle size ($K_{eff}$ values up to two orders of magnitude larger than $K_v$ have been reported for nanoparticles [11]), and, accordingly, the coercive field also decreases with increasing particle size (values up to 3000 Oe have been reported for 6 nm nanoparticles [12]). Thus, not only the order of magnitude of $H_c$ but its size dependence is anomalous in the present oleic acid-coated magnetite nanoparticles: $H_c$ increases with increasing particle size most probably due to shape anisotropy associated with the faceted-growth of the particles [4], as suggested in Figure 1. Finally, an estimate of $K_{eff}$ may be obtained by comparing the mean energy of anisotropy $K_{eff}V$ to the thermal energy $k_B<T_B>$, with $k_B$ the Boltzmann constant and $<T_B>$ the mean blocking temperature. For the typical time window of SQUID measurements, $K_{eff}V \approx 25k_B<T_B>$ [2,7]. For a log-normal distribution of particle sizes, $<T_B>$ and $T_{max}$ are related as $T_{max} \approx 2<T_B>$ [13], yielding a rough estimate $K_{eff} \approx 5·10^5$ erg/cm$^3$ for the 6 nm particles, which is indeed within the order of magnitude of the bulk anisotropy constant. This result supports the idea that the oleic acid molecules bonded covalently to the nanoparticles are able to reduce the surface spin disorder and the anisotropy is dominated by the volume contribution, such that as the diameter increases, the effective anisotropy increases due to the samples consisting of a mixture of diamond-, cubic- and triangular-shaped particles.

Preliminary x-ray photoelectron spectroscopy experiments (not shown) evidence that the binding energies of the Fe 2p core level approach to that of bulk magnetite, so that the new $O^{2-}$ surface ligands partially reconstruct the crystal field of the surface Fe cations. However, the Fe 2p core level still appears at about 1 eV below the binding energy for bulk magnetite, suggesting that the Fe-O distances at the particle surface are different from those in the bulk.

In conclusion, magnetite nanoparticles coated with oleic acid molecules bonded covalently to the surface and within the range 6-20 nm in size, show the saturation magnetization of bulk samples. Although the approach to saturation is different in both cases, the nanoparticles also show coercive fields similar to the bulk case. These findings may be of relevance in biomedical applications since they may reduce the strength of the magnetic field required to obtain a high value of the magnetisation and they open the question of whether $M_s$ above the bulk value may be obtained by taking advantage of the orbital contribution to the magnetic moment, for example, due to the interplay of the surface cations in the particles and the ligands in the surfactant [14-15].

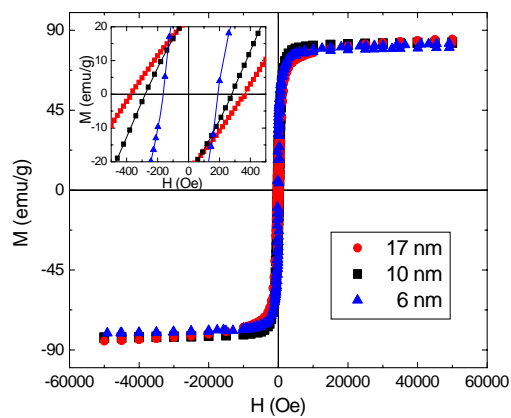

Fig. 3. Hysteresis loops at T = 5 K for magnetite nanoparticles surfaced with oleic acid, as a function of the particle size. Inset: Low magnetic field region of the hysteresis loop.

## Acknowledgments

The financial support of the Spanish MEC through the projects NAN2004-08805-CO4-02 and NAN2004-08805-CO4-01 is largely recognized.